\documentclass[review]{elsarticle}
\usepackage{algorithm}
\usepackage{algorithmic}
\usepackage{url}
\usepackage{multirow} 
\usepackage{xcolor}
\usepackage{cases}
\usepackage{color}
\usepackage{epsfig}
\begin{document}
\begin{frontmatter}

\title{Improving link prediction in complex networks by adaptively exploiting multiple structural features of networks}
\author{Chuang Ma$^{a}$}
\author{Zhong-Kui Bao $^{a}$}
\author{Hai-Feng Zhang\corref{mycorrespondingauthor}$^{a,b,c}$}\cortext[mycorrespondingauthor]{Corresponding author}
\ead{haifengzhang1978@gmail.com}

\address
{$^{a}$ School of Mathematical Science, Anhui University, Hefei
230601, China }
\address{$^{b}$ Center of Information Support \&Assurance Technology, Anhui University, Hefei, 230601, China}
\address {$^{c}$ Department of Communication Engineering, North University of China, Taiyuan, Shan'xi 030051, China}

\begin{abstract}

So far, many network-structure-based link prediction methods have been proposed. However, these methods only highlight one or two structural features of networks, and then use the methods to predict missing links in different networks. The performances of these existing methods are not always satisfied in all cases since each network has its unique underlying structural features. In this paper, by analyzing different real networks, we find that the structural features of different networks are remarkably different. In particular, even in the same network, their inner structural features are utterly different. Therefore, more structural features should be considered. However, owing to the remarkably different structural features, the contributions of different features are hard to be given in advance. Inspired by these facts, an \emph{adaptive} fusion model regarding link prediction is proposed to incorporate multiple structural features. In the model, a logistic function combing multiple structural features is defined, then the weight of each feature in the logistic function is \emph{adaptively} determined by exploiting the known structure information. Last, we use the ``learnt'' logistic function to predict the connection probabilities of missing links. According to our experimental results, we find that the performance of our adaptive fusion model is better than many similarity indices.
\begin{keyword}
 Complex networks \sep Link prediction\sep Adaptive fusion model\sep Logistic regression\sep Multiple structural features.


\end{keyword}
\end{abstract}

\date{}
\end{frontmatter}

\section{Introduction} \label{sec:intro}

The problem of link prediction in complex networks has been paid much attention in recent years. On the one hand, link prediction problem offers one possible way to understand the formation of networks. On the other hand, link prediction problem has wide range of applications, such as finding promising candidate friends in online social networks~\cite{scellato2011exploiting}, exploring possible protein-to-protein interactions~\cite{cannistraci2013link}, reconstructing airline network~\cite{guimera2009missing}, providing personalized recommendations in E-commerce systems~\cite{lu2011link,lu2015toward,lu2009similarity,zhang2016measuring,liu2015local}.

Though attribute-based algorithms have been proposed from computer science community~\cite{sarukkai2000link,popescul2003statistical}, the lack of the entity's attribute information may restrict the applications of these algorithms.  For instance, the user's personal information in social networks is hard to be obtained owing to the privacy preservation~\cite{liu2015local}. Recently, network-structure-based link prediction has become a flourishing field since the structural features of networks are easier to be obtained and the latter method also considers the structural features of networks, such as, the hierarchical organization~\cite{clauset2008hierarchical}, local-community-paradigm~\cite{cannistraci2013link}, clustering~\cite{feng2012link} and weak ties~\cite{lu2011link}. Along this line, many similarity-based indices have been proposed, and they are generally divided into two categories: local similarity indices and global similarity indices~\cite{leicht2006vertex,tong2006fast}. Since local similarity indices have the advantages of low computational complexity, simple implementation, good performance and so forth, problem on how to utilize the structural features to design an effective local similarity index has attracted much attention. For instance, common neighbors (CN) index~~\cite{newman2001clustering} assume that nodes with more common neighbors are more likely to be connected. Adamic-Adar (AA)~\cite{adamic2003friends} and resource allocation (RA)~\cite{zhou2009predicting} indices utilize the feature that the contributions from the high-degree neighbors are smaller than the low-degree neighbors. Preferential attachment (PA) index implies that high degree nodes prefer to connect each other~\cite{newman2001clustering}. Cannistraci \emph{et al.} proposed a local community
paradigm (LCP) index by taking
into account the local community feature~\cite{cannistraci2013link}.  We also proposed a friend recommendation index by utilizing the weak clique feature in networks~\cite{chuang2016playing}.

From the above descriptions, one can see that these indices were proposed by exploiting one or two structural features of networks, and then use such an index to implement link prediction to all networks~\cite{ding2016prediction,lu2011link,li2015node}. As a result, these methods imply an assumption that one considered structural feature dominates in  all networks. Obviously, the performances of these methods are discounted if the assumption is questionable. In this paper, by analyzing many real networks, we find that different networks have their inner structural features. Moreover, even in a given network, the structural features in different parts are also dramatically different. Therefore new methods which can combine multiple structural features should be proposed. In doing so, Zhu \emph{et~al.} recently have made a meaningful attempt to incorporate the multiple structural features into link prediction from the perspective of information theory; however, one parameter in their model should be given in advance~\cite{zhu2015information}.
In this paper, we propose an \emph{adaptive} fusion model with respect to link prediction to incorporate multiple structural features. In the model, we exploit the inner structural features of networks by using the logistical regression analysis, where the weight of each structural feature is \emph{adaptively} determined by the known information of the structures. By considering different cases, our experimental results indicate the good performance of our adaptive link prediction method.

%
%
%
%

\section{Preliminaries } \label{sec2}

\subsection {The problem description and evaluation metrics}
Consider an undirected network $G(V, E)$, where $V$  and $E$
are the node set and link set, respectively. For a network containing $N$ nodes, the universal possible link set, denoted by $U$, containing all $\frac{N(N-1)}{2}$ possible links. Each pair of nodes $(v_i, v_j)$ can obtain a score $S(v_i,v_j)$ according to a defined similarity index.
A higher score means a higher connection probability between $(v_i,v_j)$, and
vice versa. Since $G$ is undirected, the score is supposed to
be symmetric, that is $S(v_i,v_j) = S(v_j,v_i)$. All the nonexistent links are
sorted in a descending order according to their scores, and
the links at the top are most likely to exist~\cite{zhou2009predicting,liu2010link}.

To test the prediction accuracy of each index, the link set $E$ is randomly divided into two parts: training set $E^T$, which is supposed to be the
known information, while testing set $E^P$ is used for
testing and no information therein is allowed to be used for
prediction. As a result, $E=E^T \cup E^P$ and
$E^T\cap E^P=\emptyset$. As in previous literatures, the training set $E^T$ always contains 90\% of links in this work,
and the remaining 10\% of links constitute the testing set.  All results are averaged over
50 independent implementations.

Meanwhile, two standard metrics are used to quantify the performances of the algorithms: AUC and Precision~\cite{lu2011link}.
Area under curve (AUC) can be
interpreted as the probability that a randomly chosen missing
link (a link in $E^P$) is given a higher score than a randomly
chosen nonexistent link (a link in $U-E$).  When implementing, among $n$
independent comparisons, if there are $n'$ times where the missing
link has a higher score and $n''$ times where the two have the same score,
AUC can be written as follow~\cite{lu2011link}:
\begin{eqnarray}\label{1}
AUC=\frac{n'+0.5n''}{n}.
\end{eqnarray}

Precision is the ratio
of the number of missing links predicted correctly within
those top-$L$ ranked links to $L$. If $m$ links are correctly predicted, then Precision can be calculated as~\cite{lu2011link}:
\begin{eqnarray}\label{2}
Precision=\frac{m}{L}.
\end{eqnarray}
\subsection{Local similarity indices}

Let $\mathbf{A}$ be the adjacency matrix of the network, $\Gamma(v_i)$ be the neighbor set of node $v_i$, $|.|$ be the cardinality of the set, and $k(v_i)$ be the degree of node $v_i$. jiThere are many structural feature induced similarity indices, here we consider several local similarity indices which are used in this paper.

\textbf{CN index:} the CN index assumes that two nodes sharing more common neighbors are more likely to be connected (we also call CN feature to emphasize that the CN index is proposed to exploit the CN feature. The following indices are also called in the same fashion),
\begin{eqnarray}\label{3}
S^{CN}(v_i,v_j)=|\Gamma(v_i)\cap\Gamma(v_j)|.
\end{eqnarray}

\textbf{PA index:} the PA index emphasizes that the connection probability of a pair of nodes is proportional to their degrees'
product. Namely,
\begin{eqnarray}\label{4}
S^{PA}(v_i,v_j)=k(v_i)\cdot k(v_j).
\end{eqnarray}

\textbf{AA index:} the AA index depresses the contribution of the high-degree common neighbors,
\begin{eqnarray}\label{5}
S^{AA}(v_i,v_j)=\sum_{v_l\in\Gamma(v_i)\cap\Gamma(v_j)}\frac{1}{\lg(k(v_l))}.
\end{eqnarray}

\textbf{RA index:} the RA index is similar to the AA index, but which is motivated by the idea of resource allocation.

\begin{eqnarray}\label{6}
S^{RA}(v_i,v_j)=\sum_{v_l\in\Gamma(v_i)\cap\Gamma(v_j)}\frac{1}{k(v_l)}.
\end{eqnarray}

\textbf{LP index:} the LP index considers the tradeoff of accuracy and computational complexity,
\begin{eqnarray}\label{LP}
S^{LP}(v_i,v_j)=\mathbf{A}^2(v_i,v_j)+\alpha\mathbf{A}^3(v_i,v_j).
\end{eqnarray}
 $\mathbf{A}^2(v_i,v_j)$ and $\mathbf{A}^3(v_i,v_j)$ are the number of different paths with length 2 and length 3, respectively. $\alpha$ is a free parameter, as suggested in Ref.~\cite{lu2011link}, we set $\alpha=0.001$ to discount the impact of longer paths.

\textbf{DD index:} the DD index highlights the connection probability between a pair of nodes is related to their degree difference.
\begin{eqnarray}\label{7}
S^{DD}(v_i,v_j)=|k(v_i)-k(v_j)|,
\end{eqnarray}
$|.|$ in Eq.(\ref{7}) is the \emph{absolute value sign} rather than the cardinality of the set. Here we want to address the reason why we introduce the DD index in Eq.~(\ref{7}).  Given that  many real networks are assortative or disassortative. Owing to the different mixing patterns, the connection probability of a pair of nodes may increase or decrease with their degree difference. The logistic regression can \emph{adaptively} determine whether the connection probability increases or decreases with the degree difference. That is to say, the degree-degree correlation feature is adaptively considered in this index.

\textbf{NSI index:} in Ref.~\cite{zhu2015information}, Zhu \emph{et~al.} design a neighbor set information (NSI) index based on the information-theoretic model, in which the contributions of different structural features to link prediction are measured in the value of self-information. The connection probability between a pair of nodes is defined as:

\begin{eqnarray}\label{nsi}
{S^{NSI}}\left( {{v_i},{v_j}} \right) =  - I\left( {L_{{v_i}{v_j}}^1|{O_{{v_i}{v_j}}}} \right) - \lambda I\left( {L_{{v_i}{v_j}}^1|{P_{{v_i}{v_j}}}} \right),
\end{eqnarray}
here $I(L^1_{v_iv_j}|\omega)$ denotes the conditional self-information of the event that a pair of nodes $(v_i, v_j)$ is connected (i.e., $L^1_{v_iv_j}$) when a feature variable $\omega$ is known.

Based on the assumption that two persons are more likely to be friends if they have many common friends, or if their friends are
also mutual friends, two structural features are considered in Eq.~(\ref{nsi})---common neighbors between a pair of  nodes $(v_i, v_j)$, i.e., ${O_{{v_i}{v_j}}} = \left\{ {z:z \in \Gamma \left( {{v_i}} \right) \cap \Gamma \left( {{v_j}} \right)} \right\}$; and the links across
two neighboring sets of $v_i$ and $v_j$, which is defined as: ${P_{{v_i}{v_j}}}$ = $\{ {{l_{xy}}:{l_{xy}} \in E,x \in \Gamma \left( {{v_i}} \right),y \in \Gamma \left( {{v_j}} \right)} \}$. In the paper, the parameter $\lambda$ adjusting the contributions of the two feature is mainly set as 0.1. NSI index can be viewed as a \emph{non-adaptive} fusion model since it includes two similarity indices simultaneously but their contributions are fixed.


Our algorithm is implemented on sixteen real networks, which are drawn from different fields, including: (1) C. elegans-The neural network of the nematode worm C. elegans~\cite{watts1998collective}; (2) Email-e-mail network of University at Rovira i Virgili, URV~\cite{guimera2003self};
(3) FWEW-A 66 component budget of the carbon exchanges occurring during the wet and dry seasons in the graminoid ecosystem of South Florid~\cite{ulanowicz1998network}; (4) FWFW-A food web in Florida Bay during the rainy season~\cite{ulanowicz1998network};(5) TAP-yeast protein-protein binding network generated by tandem affinity purification experiments~\cite{gavin2002functional};(6) Power-An electrical power grid of the western US~\cite{watts1998collective}; (7) SciMet-A network of articles from or citing Scientometrics\cite{pajek_datasets}; (8) Yeast-A protein-protein interaction network in budding yeast~\cite{bu2003topological}; (9) PB-A network of the US political blogs~\cite{reese2007mapping}; (10) Facebook-Slavo Zitnik¡¯s friendship network in Facebook~\cite{blagus2012self}; (11) NS-A coauthorship network of scientists working on network theory and experiment~\cite{von2002comparative}; (12) Jazz-A collaboration network of jazz musicians~\cite{gleiser2003community}; (13) Router-A symmetrized snapshot of the structure of the Internet at the level of autonomous systems~\cite{spring2004measuring}; (14) USAir-The US Air transportation system~\cite{lu2011link}; (15) PGP-an encrypted communication
network~\cite{boguna2004models}; (16) Astro-phys-collaboration network of astrophysics scientists~\cite{newman2001structure}. Basic structural features of these networks are summarized in table~\ref{table1}.
\begin{table}
\centering
\small
\caption {The basic topological features of sixteen example networks. $N$ and $M$ are the total numbers of nodes and links,
respectively. $C$ and $r$ are the clustering coefficient and the assortative coefficient, respectively. $H$ is the
degree heterogeneity, defined as $H=\frac{\langle k^2\rangle}{\langle k\rangle^2}$, where $\langle k\rangle$ denotes
the average degree~\cite{newman2010networks}.}
\begin{tabular}{c|c|c|c|c|c}
\hline
Network &N &M  &C &r &H\\
\hline
C.elegans &297 &2148  &0.308&-0.163&1.801\\ \hline
Email &1133 &5451  &0.254&0.078&1.942\\ \hline

FWEW &69 &880  &0.552 &-0.298 &1.275\\ \hline
FWFW &128 &2075  &0.335 &-0.112 &1.237 \\  \hline

Tap &1373 &6833  &0.557 &0.579 &1.644 \\ \hline
Power &4941 &6594  &0.107 &0.003 &1.45 \\ \hline

SciMet &3084 &10399  &0.175 &-0.033 &2.78 \\ \hline
Yeast &2375 &11693  &0.388 &0.454 &3,476 \\ \hline

PB &1222 &16724  &0.36 &-0.221 &2.971 \\ \hline
Facebook &334 &2218  &0.473 &0.247 &1.615 \\ \hline

NS &1589 &2742 &0.791 &0.462 &2.011\\ \hline
Jazz &198 &2742  &0.633 &0.02 &1.395 \\ \hline

Router &5022 &6258  &0.033 &-0.138 &5.503 \\ \hline
USAir &332 &2126  &0.749 &-0.208 &3.464 \\ \hline

PGP &10680 &24316  &0.44 &0.238 &4.147 \\ \hline
Astro-phys &16706 &121251  &0.695 &0.236 &3.095 \\ \hline

\end{tabular}\label{table1}
\end{table}

\section{Unique structural features} \label{sec2}

To validate that each network has its unique structural feature, top-$|E|$ ranked links predicted by the index are chosen. In detail, the similarity score for each possible pair of nodes is calculated based on one defined index, then choose the $|E|$ pairs with the top values as the predicted links, and they are labelled as $\tilde{E}$. Since the predicted links in set $\tilde{E}$ may not correspond to the real existent links (i.e., $E$) in the network, we define the matching score $\sigma$ as:
\begin{eqnarray}\label{8}
\sigma=\frac{|E\bigcap\tilde{ E}|}{|E|}.
\end{eqnarray}
A larger value of $\sigma$ indicates the better accuracy of the index.

The CN feature and the PA feature are chosen to demonstrate that their roles in different networks are significantly different.  According to Eq.~(\ref{8}), the values of $\sigma_{CN}$ and $\sigma_{PA}$ can be calculated, respectively. Then the difference between them is defined as:
\begin{eqnarray}\label{9}
\Delta\sigma=\sigma_{CN}-\sigma_{PA}.
\end{eqnarray}

\begin{figure}
\begin{center}
\includegraphics[width=5in]{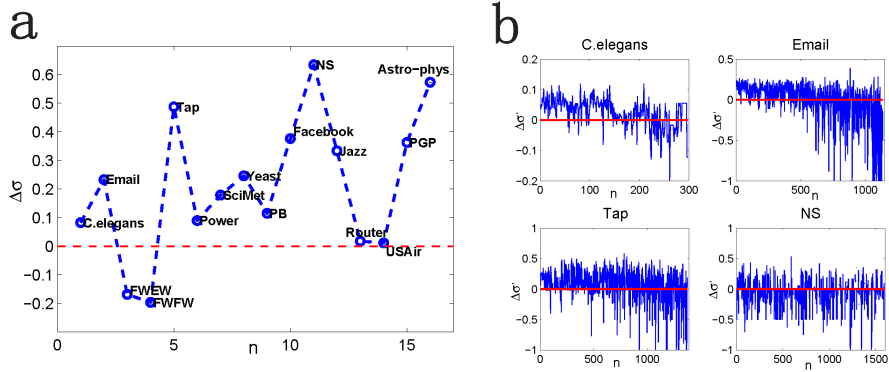}
\caption{(Color online) The difference of matching score between CN feature and PA feature. (a) In sixteen real networks, the differences $\Delta\sigma$ between CN feature and the PA feature are compared. (b) for four real networks, the difference $\Delta\sigma'$ between CN feature and PA feature in each module is plotted.}
\label{fig1}
\end{center}
\end{figure}

The values of $\Delta\sigma$ in sixteen real networks are shown in Fig.~\ref{fig1}(a). Some interesting phenomena can be observed: For FWEW and FWFW networks, the role of the PA feature is obviously superior to that of the CN feature. For Router and USAir networks, the roles of the two features are almost the same. However, for other twelve networks, the role of the CN feature is dominating, and the dominating degree varies over different networks. The results in Fig.~\ref{fig1}(a) confirm that the role of each structural feature in different networks is evidently different, as a result, it is not a wise choice to use one structural feature induced similarity index to implement link prediction in all networks.

For a given network, whether the different modules have the same feature. To answer this question, we here simply define the module of a node, which is the subgraph containing the node itself, its nearest and the next-nearest neighbors, and their inner links. Let $M(V', E')$ be a module of one network, the number of inner edges is $|E'|$.  We choose the top-$|E'|$  predicted edges in the module and labeled as $\tilde{E}'$.  Similar to Eq.~(\ref{8}) and Eq.~(\ref{9}), we define
\begin{eqnarray}\label{10}
\sigma'=\frac{|E'\bigcap \tilde{E}'|}{|E'|}.
\end{eqnarray}
In addition, the difference between $\sigma'_{CN}$ and $\sigma'_{PA}$  is denoted as:
\begin{eqnarray}\label{11}
\Delta\sigma'=\sigma'_{CN}-\sigma'_{PA}.
\end{eqnarray}
Take four real networks as examples, Fig.~\ref{fig1}(b) indicates that the structure feature in each module is also totally different.
\section{Adaptive fusion model base on logistic regression} \label{sec3}

The results in the above section indicate that:  the roles of different structural features in networks are totally different; 2) even in the same network, the role of each structural feature in different modules is also different. Therefore, it is not wise to use one structural feature induced index to predict missing links in different networks. A reasonable method should be able to exploit the inner structural features themselves and then use these structural information to predict missing links. In addition, since the contribution of each structural feature varies with networks and modules, its contribution should be \emph{adaptively} determined rather than given in advance. In doing so, an adaptive fusion model based on logistic regression (for the sake of simplicity, the similarity index proposed by this method is labelled as: LR index) to predict missing links is proposed.

Let $F_l$ be the $l$th structural feature, and $M_k$ be the $k$th module in the network. $S^{F_l}_{M_k}(v_i,v_j)$ is the similarity score induced by feature $F_l$ for a pair of $(v_i, v_j)$ in the module $M_k$. Specially, $S^{F_l}_{M_k}(v_i,v_i)\equiv0$ since there is no any self-loops in networks.

 For a given module $M_k$, the connection probability of $(v_i, v_j)$ incorporating different features can be generally written as:
\begin{eqnarray}\label{12}
P_{M_k}(v_i,v_j)=f(S^{F_1}_{M_k}(v_i,v_j),S^{F_2}_{M_k}(v_i,v_j),\cdots,S^{F_L}_{M_k}(v_i,v_j)).
\end{eqnarray}
It is worth noting that a pair of nodes $(v_i, v_j)$ can coexist in different modules, so the final connection probability  of $(v_i, v_j)$ is defined as the maximum:
\begin{eqnarray}\label{13}
P(v_i,v_j)=\max\{P_{M_1}(v_i,v_j),\cdots,P_{M_k}(v_i,v_j)\}.
\end{eqnarray}
 Eqs.~(\ref{12}) and (\ref{13}) just provide a general framework to calculate the connection probability (or score), however, how to choose a proper function form in Eq.~(\ref{12}) is an important issue. More importantly, the weight of each feature in different modules is also different. We therefore use the logistic regression method to overcome such a problem~\cite{hilbe2009logistic}, in which we use the known information in the module to \emph{adaptively} ``learn'' the weights of different structural features. Namely, the probability of a pair of node $(v_i, v_j)$ in module $M_k$ is determined as:
\begin{eqnarray}\label{15}
P_{M_k}(v_i,v_j)=\frac{1}{1+e^{-(\beta_0+\sum^{L}_{l=1}\beta_l S^{F_l}_{M_k}(v_i,v_j))}}.
\end{eqnarray}
From Eq.~(\ref{15}) one can find that the feature $F_i$ ($i=1,2,\cdots,L$) is favored when $\beta_i>0$, on the contrary, the feature $F_i$ is depressed when $\beta_i<0$. The values of $\beta_0, \beta_1,\cdots, \beta_L$ can be obtained by using the known information of the existent links in the module.

For a network containing $N$ nodes, each node and the nodes who are near to it forms a defined module. As a result, the network gives rise to $N$ modules and different modules can have many overlapping nodes. And because of this, the final connection probability is the maximal value of connection probability in different modules (see Eq.~(\ref{13})). We consider three scenarios of modules to check our algorithm:

(1): each node combining its nearest neighbors form a module (we use $LR_{1}$ to denote the similarity index based on this case);

(2): each node combining its nearest and next-nearest neighbors form a module (we use $LR_{2}$ to denote the index based on this case).

(3): when networks are very sparse, the size of the module defined in case (1) may be very small, i.e., the known information is too little to fit the  parameters in Eq.~(\ref{15}). On the contrary, the size of the module defined in case (2) may be too large, which cannot distinguish the difference of the inner structural features. Therefore, a mixed module is defined as: module is defined as the case (1) if its size is larger than 10, otherwise, we defined the module as the case (2). (we use $LR_{m}$ to denote the index based on this case).

The main steps of our method are summarized as:

\textbf{Step 1:} For a network with $N$ nodes, the $N$ modules are obtained based on $LR_{1}$, $LR_{2}$, or $LR_{m}$;\\

\textbf{Step 2:} For each module, we calculate some values according to the known information in the module, such as the similarity score of each pair based on the CN index, PA index, DD index, and so forth;\\

\textbf{Step 3:} Select several typical features (such as CN feature, PA feature or DD feature) in the logistic function Eq.~(\ref{15}). Then the values of $\beta_i$ in Eq.~(\ref{15}) are obtained such that the connection probabilities of the existent links (i.e., $\mathbf{A}(v_i,v_j)=1$) are the largest. In this way, the logistic function in each module is determined, and the connection probability of each pair of nodes (including nonexistent pairs) can be calculated;\\

\textbf{Step 4:} According to Eq.~(\ref{13}), the similarity score for each  pair of nodes is finally given. Algorithm 1 presents the detailed procedure of adaptive fusion model regarding link prediction.

\begin{algorithm}
\renewcommand{\algorithmicrequire}{\textbf{Input:}}
\renewcommand{\algorithmicensure}{\textbf{Output:}}

\caption{Algorithm of the adaptive fusion model based on the logistic regression.}
\label{algo:algorithm1}
\begin{algorithmic}[1]

\REQUIRE
{Network $G=(V,E)$}
\ENSURE
{Probability matrix $P$}
\STATE
$P \leftarrow zero \, matrix$

\FOR{${\bf each} \ v_k \in V $}
    \STATE
    Find the module ${M_k} $ based on ${LR_1} $, ${LR_m} $  or ${LR_2} $ by node ${v_k} $
    \STATE
    /* The module ${M_k} $ is a set of nodes*/
    \STATE
    $n \leftarrow 1$
    \FOR{${\bf each} \ v_i , v_j(i\geq j)\in M_k $}
        \STATE
        $y(n) \leftarrow \mathbf{A}(v_i , v_j)$
        \STATE
        /*$A$ is adjacency matrix of network $G$*/
        \STATE
        $x_{1}(n) \leftarrow S_{M_k}^{F_1}(v_i,v_j),\cdot\cdot\cdot, x_L(n) \leftarrow S_{M_k}^{F_L}(v_i,v_j)$

        \STATE
        /* $S_{M_k}^{F_l}(v_i,v_j)$ is the similarity score induced by feature $F_l$ for a pair of $(v_i,v_j)$ in the module $M_k$*/

        \STATE
        $n \leftarrow n+1$
    \ENDFOR
    \STATE
    Compute $\beta_0,\beta_1,\cdot\cdot\cdot,\beta_L$ by $y = \frac{1}{{1 + {e^{ - ({\beta _0} + \sum\nolimits_{l = 1}^L {{\beta _l}{x_l}} )}}}}$ to fit the values $y,x_1,\cdot\cdot\cdot,x_L$
   \FOR{${\bf each} \ v_i , v_j\in M_k $}

        \STATE
        Compute $P_{M_k}(v_i,v_j)$ by using Equation (\ref{13})
        \STATE
        $P(v_i,v_j)=max\{ P(v_i,v_j),P_{M_k}(v_i,v_j)   \}$

    \ENDFOR

\ENDFOR
\STATE \textbf{return} $P$
\end{algorithmic}
\end{algorithm}

\begin{figure}
\begin{center}
\includegraphics[width=5in]{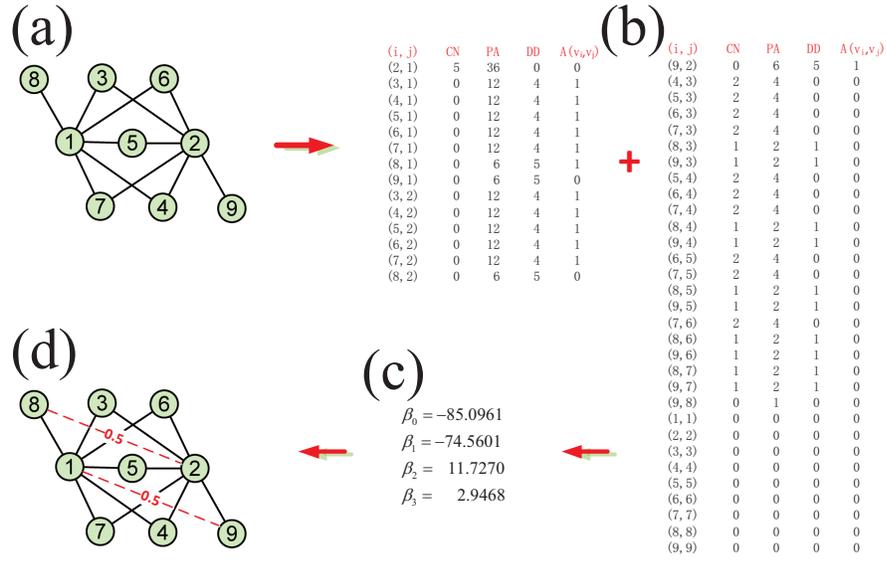}
\caption{(Color online) An illustration is given to explain how to determine the values of parameters in Eq.~(\ref{15}). (a) a toy network with 9 nodes. We assume that the information appeared in this network is known, then we need to predict the connection probability of each nonexistent link, (b) the values of $S^{CN}$, $S^{PA}$,$S^{DD}$ and $\mathbf{A}(v_i,v_j)$ regarding each pair of nodes $(v_i, v_j)$ are calculated, (c) obtain the values of $\beta_0$, $\beta_1$, $\beta_2$ and $\beta_3$ by using Eq.~(\ref{15}) to fit the values in (b), (d) calculate the connection probability of each nonexistent link, and all of them are equal to zero except for $P(v_2,v_8)=0.5$ and $P(v_1,v_9)=0.5$ (marked by red dash lines).}
\label{fig0}
\end{center}
\end{figure}

To clarify our method, an illustration in Fig.~\ref{fig0} is given to explain how to determine the values of parameters in Eq.~(\ref{15}). For a toy network (see Fig.~\ref{fig0}(a)), according to the known information of the network, for a pair node $(v_i, v_j)$,   $\mathbf{A}(v_i,v_j)=1$ when they are connected, or $\mathbf{A}(v_i,v_j)=0$ otherwise. Meanwhile, the values of $S^{CN}$, $S^{PA}$ and $S^{DD}$ regarding to each pair of nodes $(v_i, v_j)$ can be calculated (Fig.~\ref{fig0}(b)). Then the values of $\beta_0$, $\beta_1$, $\beta_2$ and $\beta_3$ in Eq.~(\ref{15}) can be \emph{adaptively} determined by using Eq.~(\ref{15}) to fit the values in Fig.~\ref{fig0}(b) (see Fig.~\ref{fig0}(c)).  Once the values of these parameters are given, the connection probability of each nonexistent link can be calculated. the results indicate that all of them are equal to zero except for $P(v_2,v_8)=0.5$ and $P(v_1,v_9)=0.5$ (see Fig.~\ref{fig0}(d)).

\section{Main results and analysis} \label{sec4}

At first, we consider a scenario incorporating CN, PA and DD features into the adaptive fusion model. Similar to the steps in Fig.~\ref{fig0}, once the module of each node is defined, the parameters in Eq.~(\ref{15}) can be determined by fitting the known information within the module. Therefore, the connection probability of each pair can be easily calculated.

The values of AUC for different indices in sixteen real networks
are shown in table~\ref{table2}. By comparing each row, one can see that the performance of the logistic regression based methods ($LR_1$, $LR_2$ and $LR_m$) is generally better than other indices, or near the highest value of AUC. Even though the performances of NSI index and RA index are the best in NS network, Jazz network and USAir network, respectively, the performance of our LR index is very close to them.  Moreover, the performances of the $LR_m$ and $LR_2$ indice are more remarkable (the highest values of AUC are emphasized by bold font).

\begin{figure}
\begin{center}
\includegraphics[width=5in]{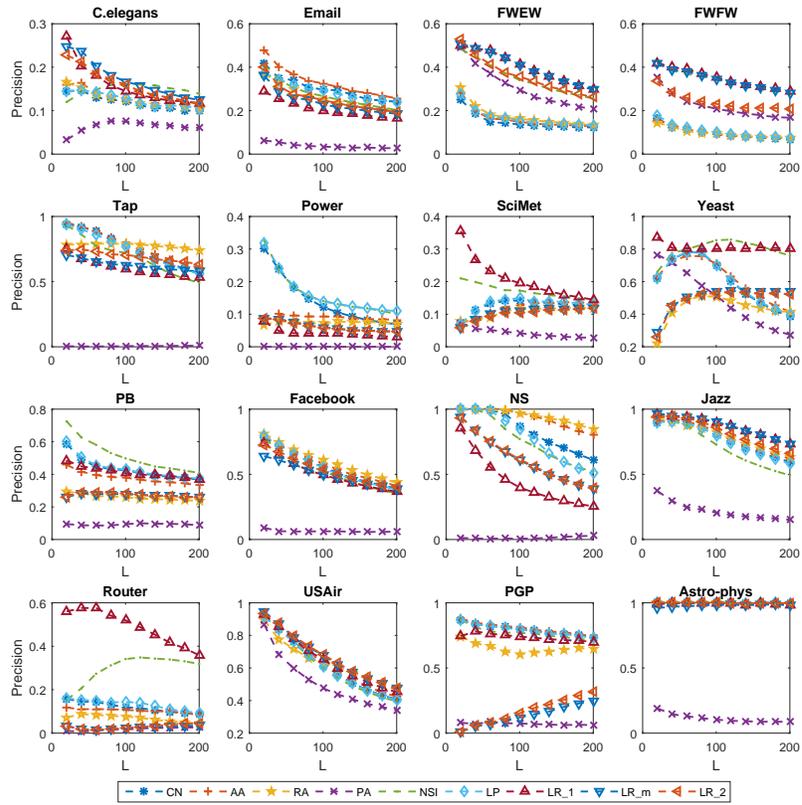}
\caption{(Color online) On sixteen real networks,
Precision as a function of $L$ in Eq.~(\ref{2}) is illustrated. Here the CN, PA and DD features are incorporated into the adaptive fusion model. }
\label{fig3}
\end{center}
\end{figure}

The dependence of Precision on the number of $L$ in sixteen real networks is presented in Fig.~\ref{fig3}, it demonstrates that LR index can achieve a high Precision accuracy in most networks, but the performance of PA index is the worst in most cases. Therefore, we can conclude that our method overall outperforms other indices, regardless of whether AUC metric or Precision metric.

\begin{table}
\centering
\tiny
\caption {The comparison of algorithms' accuracy quantified by AUC on sixteen networks. Here the CN, PA and DD features are incorporated into the adaptive fusion model. The highest value of AUC in each row is emphasized by bold color.}
\begin{tabular}{c|c|c|c|c|c|c|c|c|c}
\hline
Network &CN &AA  &RA &PA &NSI  &LP &LR\underline{ }1 &LR\underline{ }m &LR\underline{ }2\\
\hline
C.elegans &0.845 &0.8606 &0.8649	&0.7547	&0.8647	 &0.863	&0.8555	&\textbf{0.8879}	&0.8807
  \\ \hline
Email &0.8562	&0.8579	&0.8576	&0.8044		&0.9204 &0.919	&0.8542	&0.919	&\textbf{0.9252}
  \\ \hline

FWEW &0.691	&0.6982	&0.7053	&0.8168		&0.8539&0.7092	&\textbf{0.8781}	&0.8776	&0.8369
 \\ \hline
FWFW &0.6066	&0.609	&0.6129	&0.7342		&0.8213&0.6226	&\textbf{0.8435} &0.8423&0.7781
  \\ \hline

Tap &0.9548	&0.9554	&0.9556	&0.7247	&0.9685&0.969	&0.9527	&0.9715	&\textbf{0.9736}
\\ \hline
Power &0.6243	&0.6242	&0.6248	&0.5798		&0.6964 &0.6964	&0.6242	&0.7523	&\textbf{0.7546}
 \\ \hline

SciMet &0.7968	&0.7984	&0.7982	&0.8107		&0.9134&0.9101	&0.7968	&0.9164	&\textbf{0.9281}
 \\ \hline
Yeast &0.9125	&0.9133	&0.9135	&0.8628		&0.97 &0.9702	&0.9129	&0.9722	&\textbf{0.9741}
 \\ \hline

PB &0.925	&0.9284 &0.9293	&0.9104		&0.9431 &0.9374	&0.9347	&\textbf{0.9457}	&0.9428
\\ \hline
Facebook &0.9423 &0.9469	&0.9481	&0.7571		&0.9371 &0.9438	&0.9345	&0.9504	&\textbf{0.9523}
\\ \hline

NS &0.9908	&0.991	&0.9911	&0.6811 &\textbf{0.9974} &0.9971	&0.9892	&0.9937	&0.9917
 \\ \hline
Jazz &0.957	&0.964	&\textbf{0.9728}	&0.7711	 &0.9323 &0.9532	&0.9698	&0.9726	&0.9695
\\ \hline

Router &0.6521	&0.6523	&0.6521	&0.9553	&0.9464&0.9444	&0.6525	&0.951	&\textbf{0.9659}
 \\ \hline
USAir &0.9562	&0.9674	&\textbf{0.9737} &0.9142		&0.9472 &0.9539	&0.9564	&0.966	&0.9644
 \\ \hline

PGP &0.9419	&0.9425	&0.9421	&0.8809	&0.9741	&0.9743	&0.942	&0.9773	&\textbf{0.983}
 \\ \hline
 Astro-phys &0.9924	&0.9928	&0.9926	&0.8456	&0.9939 &\textbf{0.9952}	&0.9921	&\textbf{0.9952}	&0.9935
 \\ \hline

\end{tabular}\label{table2}
\end{table}

\begin{figure}
\begin{center}
\includegraphics[width=5in]{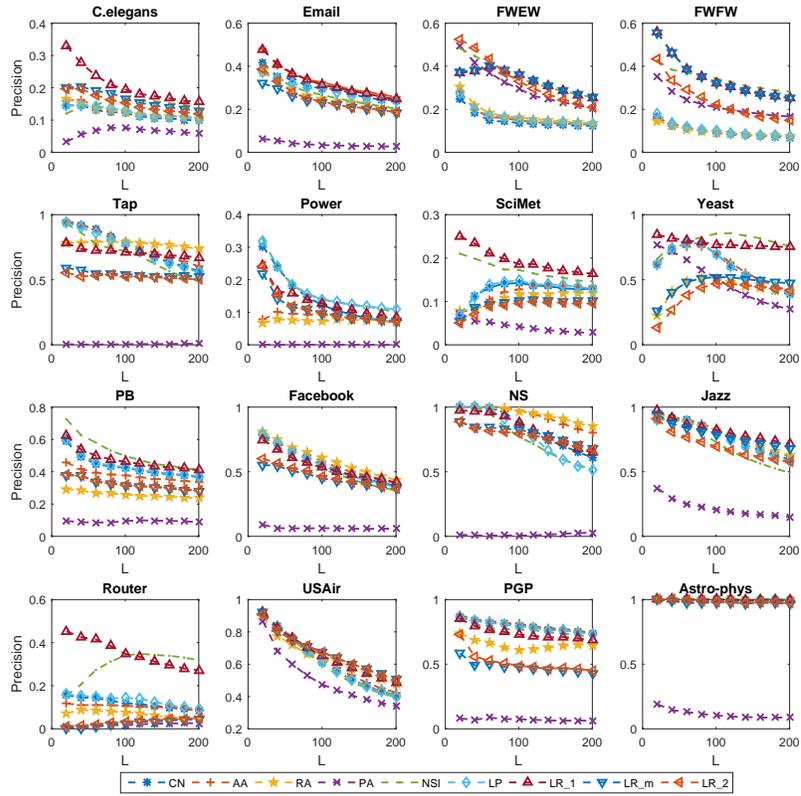}
\caption{(Color online)  On sixteen real network, the dependence of
Precision on the number of $L$ in Eq.~(\ref{2}) is plotted.  Here the CN, RA and DD features are incorporated into the adaptive fusion model. }
\label{fig4}
\end{center}
\end{figure}

To investigate whether the options of structural features play significant roles in the performance of our LR index, a new scenario combing CN, RA and DD features into the logistic regression is studied. That is to say, PA feature is replaced by RA feature. The values of AUC regarding different indices compared in sixteen real networks are summarized in table~\ref{table3}. Except for the best performance of NSI index in FWEW, FWFW and NS network, our LR index outperforms the other indices overall.
The Precision as a function of $L$ is plotted in Fig.~\ref{fig4}, the results indicate that the LR index can also guarantee the high value of Precision. Moreover, by comparing Fig.~\ref{fig3} with Fig.~\ref{fig4}, one can observe that the Precision in Fig.~\ref{fig4} is generally larger than that of in Fig.~\ref{fig3}. This is because that the performance of RA index in Precision is generally better than that of PA index.

Results in the two scenarios have confirmed that the performance of LR index is better than the other similarity indices. Therefore, we only need to select several main structural features in the adaptive fusion model, because our method is not largely sensitive to the selection of structural features. As we have stated, each network or each module has its unique structural features, many previous similarity indices only exploit one or two structural features of the networks, or the weights of different structural features are artificially given. The performance of these similarity indices is not good enough since the structural diversity of networks is not taken into account. On the contrary, our method first divides networks into different modules, and the weights of the different structural features in each modules are adaptively determined by the known structure information. That is to say, the structural diversity in each module and each network is sufficiently exploited.

\begin{table}
\centering
\tiny
\caption {The comparison of algorithms' accuracy quantified by AUC on sixteen networks. Here the CN, RA and DD features are incorporated into the adaptive fusion model. The highest value of AUC in each row is marked by bold color.}
\begin{tabular}{c|c|c|c|c|c|c|c|c|c}
\hline
Network &CN &AA  &RA &PA   &NSI  &LP &LR\underline{ }1 &LR\underline{ }m &LR\underline{ }2\\
\hline
C.elegans &0.845 &0.8606 &0.8649 &0.7547 &0.8772 &0.863	&0.8728	&\textbf{0.8965} &0.8865
  \\ \hline
Email &0.8562	&0.8579	&0.8576	&0.8044		&0.9204 &0.919	&0.8569	&0.9197	&\textbf{0.9256}
  \\ \hline

FWEW &0.691	&0.6982	&0.7053	&0.8168		&\textbf{0.8539}&0.7092	&0.8202	&0.8193	&0.7247
 \\ \hline
FWFW &0.6066	&0.609	&0.6129	&0.7342		&\textbf{0.8213}&0.6226	&0.7737 &0.7689&0.5892
  \\ \hline

Tap &0.9548	&0.9554	&0.9556	&0.7247		&0.9685&0.969	&0.9552	&0.9724	&\textbf{0.9745}
\\ \hline
Power &0.6243	&0.6242	&0.6248	&0.5798		&0.6964 &0.6964	&0.6244 &0.7525	&\textbf{0.7547}
 \\ \hline

SciMet &0.7968	&0.7984	&0.7982	&0.8107		&0.9134&0.9101	&0.7983	&0.9168	&\textbf{0.9286}
 \\ \hline
Yeast &0.9125	&0.9133	&0.9135	&0.8628	&0.97 &0.9702	&0.9142	&0.9726	&\textbf{0.9747}
 \\ \hline

PB &0.925	&0.9284 &0.9293	&0.9104		&0.9431 &0.9374	&0.9352	&\textbf{0.9455}	&0.9421
\\ \hline
Facebook &0.9423 &0.9469	&0.9481	&0.7571		&0.9371 &0.9438	&0.9464	&0.9535	&\textbf{0.9538}
\\ \hline

NS &0.9908	&0.991	&0.9911	&0.6811  &\textbf{0.9974} &0.9971	&0.9904	&0.9961	&0.9953
 \\ \hline
Jazz &0.957	&0.964	&0.9728	&0.7711 &0.9323 &0.9532	&0.9802	&\textbf{0.9804}	&0.9727
\\ \hline

Router &0.6521	&0.6523	&0.6521	&0.9553 	&0.9464&0.9444	&0.6527	&0.9515	&\textbf{0.966}
 \\ \hline
USAir &0.9562	&0.9674	&0.9737 &0.9142		&0.9472 &0.9539	&0.9687	&\textbf{0.9739}&0.9723
 \\ \hline

PGP &0.9419	&0.9425	&0.9421	&0.8809	&0.9741	&0.9743	&0.942	&0.9775	&\textbf{0.9837}
 \\ \hline
 Astro-phys &0.9924	&0.9928	&0.9926	&0.8456		&0.9939 &0.9952	&0.9928	&\textbf{0.9957}	&0.9954
 \\ \hline

\end{tabular}\label{table3}
\end{table}

\section{Conclusions} \label{sec7}
 In this paper, we first confirmed that each network or each module has its unique structural features, and it means that we cannot use one feature induced similarity index to predict missing links in all networks. Meanwhile, it is reasonable to design a link prediction algorithm in which the weight of each structural feature is \emph{adaptively} determined rather than artificially given in advance. In view of these facts, we have designed an \emph{adaptive} fusion model regarding link prediction based on the logistic regression. In the model, the weights of structural features in each module are exploited by using logistic function to fit the partial known information, i.e., the parameters in logistic function are determined. Then the connection probability of each pair of nodes is calculated. Since our fusion model sufficiently mines the known structure information and uses the information to adaptively determine the weight of different structural features, which ensures the performance of our algorithm is significantly better than other local similarity indices, regardless of whether AUC metric or Precision metric. Moreover, the proposed LR index is a local index since we only use the information of the nearest neighbors and the next-nearest neighbors, which can greatly reduce the complexity of algorithm.


\section*{Acknowledgments}

This work is supported by National Natural Science Foundation of China (61473001, 11331009), and partially supported by the Young Talent Funding of Anhui Provincial Universities (gxyqZD2017003).
\bibliographystyle{model1a-num-names}

\end{document}